\documentclass{IEEEtran}
\usepackage[noadjust]{cite}

\usepackage{amsmath,amssymb,amsfonts}
\usepackage{graphicx}
\usepackage{textcomp,nicefrac}
\usepackage{xcolor}
\usepackage[hyphens]{url} 

\def\BibTeX{{\rm B\kern-.05em{\sc i\kern-.025em b}\kern-.08em
T\kern-.1667em\lower.7ex\hbox{E}\kern-.125emX}}
\IEEEoverridecommandlockouts

\usepackage{tikz}
\usepackage{textcomp}
\usepackage{lipsum}

\newcommand\copyrighttext{%
  \footnotesize \textcopyright 
  2024 IEEE. Personal use of this material is permitted. Permission from IEEE must be obtained for all other uses, in any current or future media, including reprinting/republishing this material for advertising or promotional purposes, creating new collective works, for resale or redistribution to servers or lists, or reuse of any copyrighted component of this work in other works. 
  DOI: {10.1109/TNS.2024.3355473}}
\newcommand\copyrightnotice{%
\begin{tikzpicture}[remember picture,overlay]
\node[anchor=south,yshift=6pt] at (current page.south) {\fbox{\parbox{\dimexpr\textwidth-\fboxsep-\fboxrule\relax}{\copyrighttext}}};
\end{tikzpicture}%
}
\begin{document}
\title{Energetic particle contamination in STIX during Solar Orbiter's passage through Earth's radiation belts and an interplanetary shock}

\author{Hannah Collier, Olivier Limousin, Hualin Xiao, Arnaud Claret, Frederic Schuller, Nina Dresing, Saku Valkila, Francisco Espinosa Lara, Annamaria Fedeli, Simon Foucambert and Säm Krucker
\thanks{Solar Orbiter is a space mission of international collaboration between ESA and NASA, operated by ESA. The STIX instrument is an international collaboration between Switzerland, Poland, France, Czech Republic, Germany, Austria, Ireland, and Italy. H.C., H.X. and S.K. are supported by the Swiss National Science Foundation Grant 200021L\_189180 for STIX. N.D., A.F., and S.V. are grateful for support by the Academy of Finland (SHOCKSEE, grant No.\ 346902). F.S. is supported by the German space agency (DLR, Deutsches Zentrum für Luft- und Raumfahrt), with grant numbers 50OT1904 and 50OT2304.}
\thanks{H. Collier is with Fachhochschule Nordwestschweiz, Bahnhofstrasse 6, 5210 Windisch, Switzerland and ETH Z\"{u}rich, R\"{a}mistrasse 101, 8092 Z\"{u}rich Switzerland (e-mail: hannah.collier@fhnw.ch).}
\thanks{O. Limousin is with  Université Paris-Saclay, Université Paris Cité, CEA, CNRS, AIM (email: Olivier.LIMOUSIN@cea.fr).}
\thanks{H. Xiao is with Fachhochschule Nordwestschweiz, Bahnhofstrasse 6, 5210 Windisch, Switzerland (email: hualin.xiao@fhnw.ch).}
\thanks{A. Claret is with Université Paris-Saclay, Université Paris Cité, CEA, CNRS, AIM (email: arnaud.claret@cea.fr).}
\thanks{F. Schuller is with the Leibniz-Institut für Astrophysik Potsdam (AIP), An der Sternwarte 16, 14482 Potsdam, Germany (email: fschuller@aip.de).}
\thanks{N. Dresing is with the Department of Physics and Astronomy, University of Turku, 20014 Turku, Finland (email: nina.dresing@utu.fi).}
\thanks{S. Valkila is with the Department of Physics and Astronomy, University of Turku, 20014 Turku, Finland (email: t09saval@utu.fi).}
\thanks{F. Espinosa Lara is with the Universidad de Alcalá Space Research Group, University of Alcalá de Henarez, Spain (email: francisco.espinosal@uah.es).}
\thanks{A. Fedeli is with the Department of Physics and Astronomy, University of Turku, 20014 Turku, Finland (email: annamaria.fedeli@utu.fi).}
\thanks{S. Foucambert is with  Université Paris-Saclay, Université Paris Cité, CEA, CNRS, AIM (email: simon.foucambert@cea.fr).}
\thanks{S. Krucker is with Fachhochschule Nordwestschweiz, Bahnhofstrasse 6, 5210 Windisch, Switzerland and the Space Sciences Laboratory, University of California, 7 Gauss Way, 94720 Berkeley, USA (email: samuel.krucker@fhnw.ch).}}

\maketitle
\copyrightnotice

\begin{abstract}
The Spectrometer/Telescope for Imaging X-rays (STIX) is a hard X-ray imaging spectrometer on board the ESA and NASA heliospheric mission Solar Orbiter. STIX has been operational for three years and has observed X-ray emission from $\sim35,000$ solar flares. Throughout its lifetime, Solar Orbiter has been frequently struck by a high flux of energetic particles usually of flare origin, or from coronal mass ejection shocks. These Solar Energetic Particles (SEPs) are detected on board by the purpose-built energetic particle detector instrument suite. During SEP events, the X-ray signal is also contaminated in STIX. This work investigates the effect of these particles on the STIX instrument for two events. The first event occurred during an interplanetary shock crossing and the second event occurred when Solar Orbiter passed through Earth's radiation belts while performing a gravity assist maneuver. The induced spectra consist of tungsten fluorescence emission lines and secondary Bremsstrahlung emission produced by incident particles interacting with spacecraft components. For these two events, we identify $> 100$ keV electrons as significant contributors to the contamination via Bremsstrahlung emission and tungsten fluorescence. 
\end{abstract}


\section{Introduction}
\label{sec:introduction}
Solar Orbiter is an ESA led solar and heliospheric mission. The spacecraft is in an elliptical orbit around the Sun reaching a distance of 0.3 AU at perihelion \cite{muller2020}. The Spectrometer/Telescope for Imaging X-rays (STIX) is a hard X-ray (HXR) imaging spectrometer on board. STIX detects photons in the energy range 4-150 keV with a 1 keV resolution (at 6 keV) \cite{Krucker_2020} which  predominantly come from solar flares \cite{fletcher2011, kontar2011, krucker2008}. It is an indirect Fourier imager that uses a set of tungsten grids in front of 32 coarsely pixelated CdTe detectors \cite{Krucker_2020}. When the spacecraft is hit by a high flux of energetic particles such as during a Solar Energetic Particle (SEP) event, during a gravity assist maneuver or from Coronal Mass Ejection (CME) related shocks \cite{Antiochos1999, webb2012, sheeley1985}, the instrument detects excess HXR counts. This contributes unwanted background to the flare signal that makes the data more challenging to analyze. This work aims to understand the particle species and the corresponding mechanism(s) responsible for this contamination. This analysis is important because it is imperative for understanding the instrumental response, for performing scientific analysis, and for informing future instrument design.

One advantage of conducting this analysis with data from Solar Orbiter is that there is a complimentary suite of instruments on board. In particular, the Energetic Particle Detector (EPD) detects energetic particles with high temporal resolution over a wide energy range from suprathermal energies up to several hundreds of MeV/nucleon during these events \cite{epd_2020}. Measurements by EPD are used to compare the observed X-ray signal to the flux of energetic particles at the spacecraft.

In this work, two events are presented. One of the events occurred when the Solar Orbiter spacecraft passed through an interplanetary (IP) shock associated with a CME. The other event occurred when the spacecraft entered Earth's radiation belts \cite{Koskinen_Kilpua_book} as part of a one-off Earth Gravity Assist Maneuver (EGAM). The flux of energetic particles reaching the spacecraft during both events caused HXR counts well above instrumental background. The observed photon spectra and the spatial structure of the contamination in the two events have instructive similarities. 

\section{Observations}

\subsection{Interplanetary shock crossing} \label{subsec:IP_shock_obs}
\begin{figure*}
    \centering
    \includegraphics[width=\textwidth]{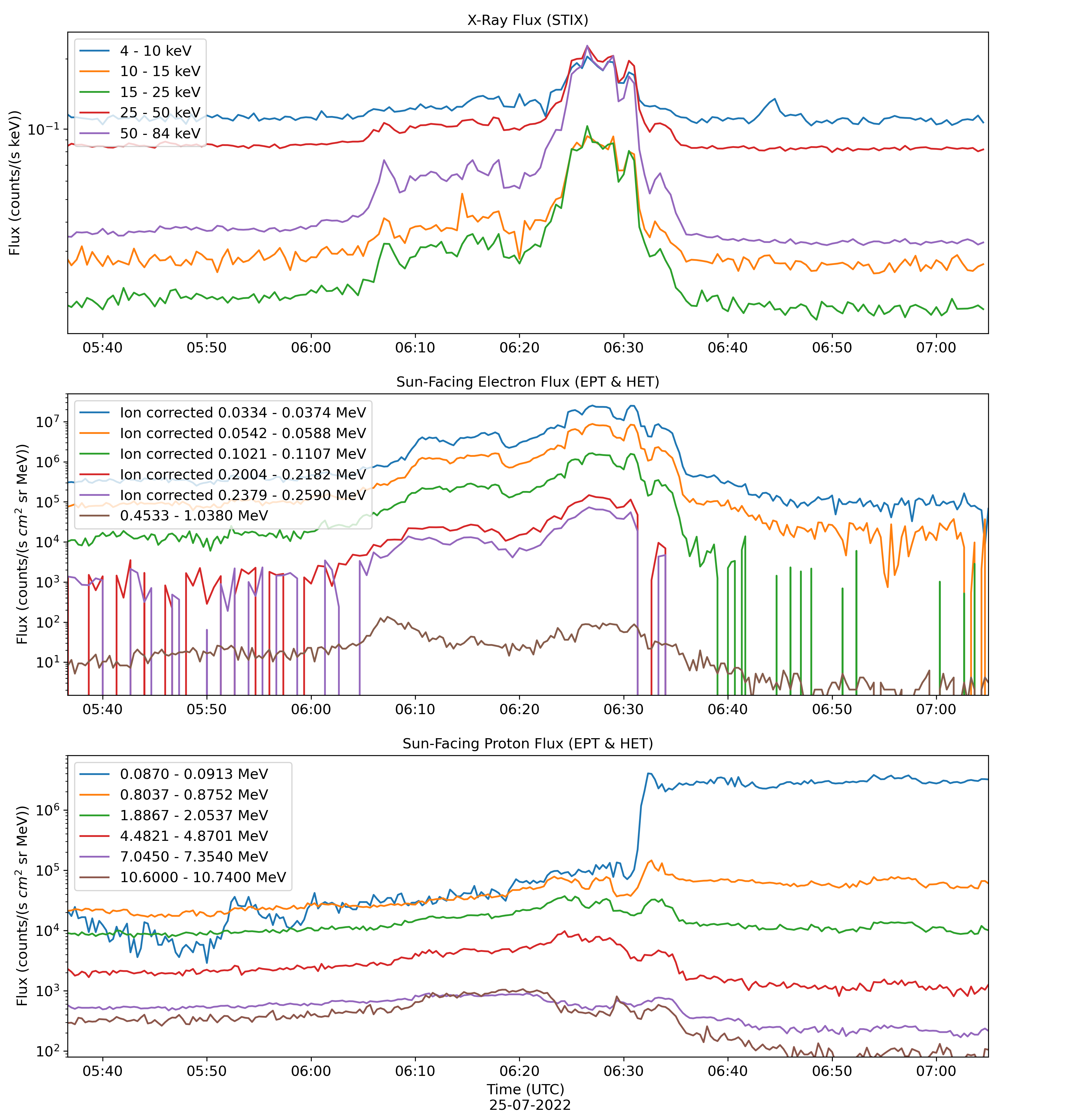}
    \caption{Energetic electrons and protons detected by the EPT and HET instruments of the EPD instrument suite compared to the observed HXR signal measured by STIX during the July 25th 2022 IP shock crossing. The particle time series have been resampled at a 20 s cadence, except for the two highest energy EPT channels shown which were resampled at 40 s. A correction for ion contamination has been applied to the EPT data. The higher energy EPT electron channels ($\gtrapprox 250$ keV) are dominated by energetic ions and thus are not shown here. Only the time profile of the HET channel that suffered least from ion contamination is shown. The HXR time profile correlates with the low energy electron EPT channels and the first HET channel. The protons display a more gradual time evolution, with a sudden sharp increase after 06:30 in the lowest energy EPT proton channels.}
    \label{fig:shock_ts}
\end{figure*}

On July 25th, 2022 Solar Orbiter passed through a CME shock. The CME occurred as a result of a solar eruption on the 23rd of July, two days previous. During this encounter, excess background counts in the HXR channels of STIX were detected due to the high flux of energetic particles. EPD directly measured these shock accelerated particles. Fig. \ref{fig:shock_ts} shows a light curve of the signal in STIX compared to the electrons and ions detected by EPD's Electron-Proton Telescope (EPT) and High Energy Telescope (HET).

By comparing the time evolution of electron and ion channels in the two EPD instruments, potential candidates for the contamination were identified. The time profiles of electrons with energies in the range 0.01-7.00 MeV and protons with energies in the range 0.05-80 MeV were analyzed. There was no increase in protons that seemed to correlate with the HXR light curves at the time of enhanced HXR signal. However, the overall shape and timing of the arrival of the low energy ($\approx 50$ keV) EPT electron population correlates with the signal observed in the STIX detectors.

The higher energy EPD electron channels ($\gtrapprox 200$ keV) are significantly contaminated by energetic ions, including those measured by EPT and HET. A first order correction has been applied to the EPT data. The correction uses the ion intensities measured by the EPT ion channels and the known response to protons of the electron channels for estimating the contamination, which is then subtracted from the electron data. It assumes that the ion measurements are dominated by protons (other ion species do not produce a noticeable contamination, but may also be present in the measured intensities leading to a slight over-correction). The HET channel which suffered least from ion contamination (0.4533-1.0383 MeV) is also shown. The time profile of these $\sim 1$ MeV electrons correlates well with the HXR signal. In particular, the peak at approximately 06:07 is also present in the HXR profiles but is not present in lower energy EPT electron channels.   

\begin{figure}
    \centering
    \includegraphics[width=\columnwidth]{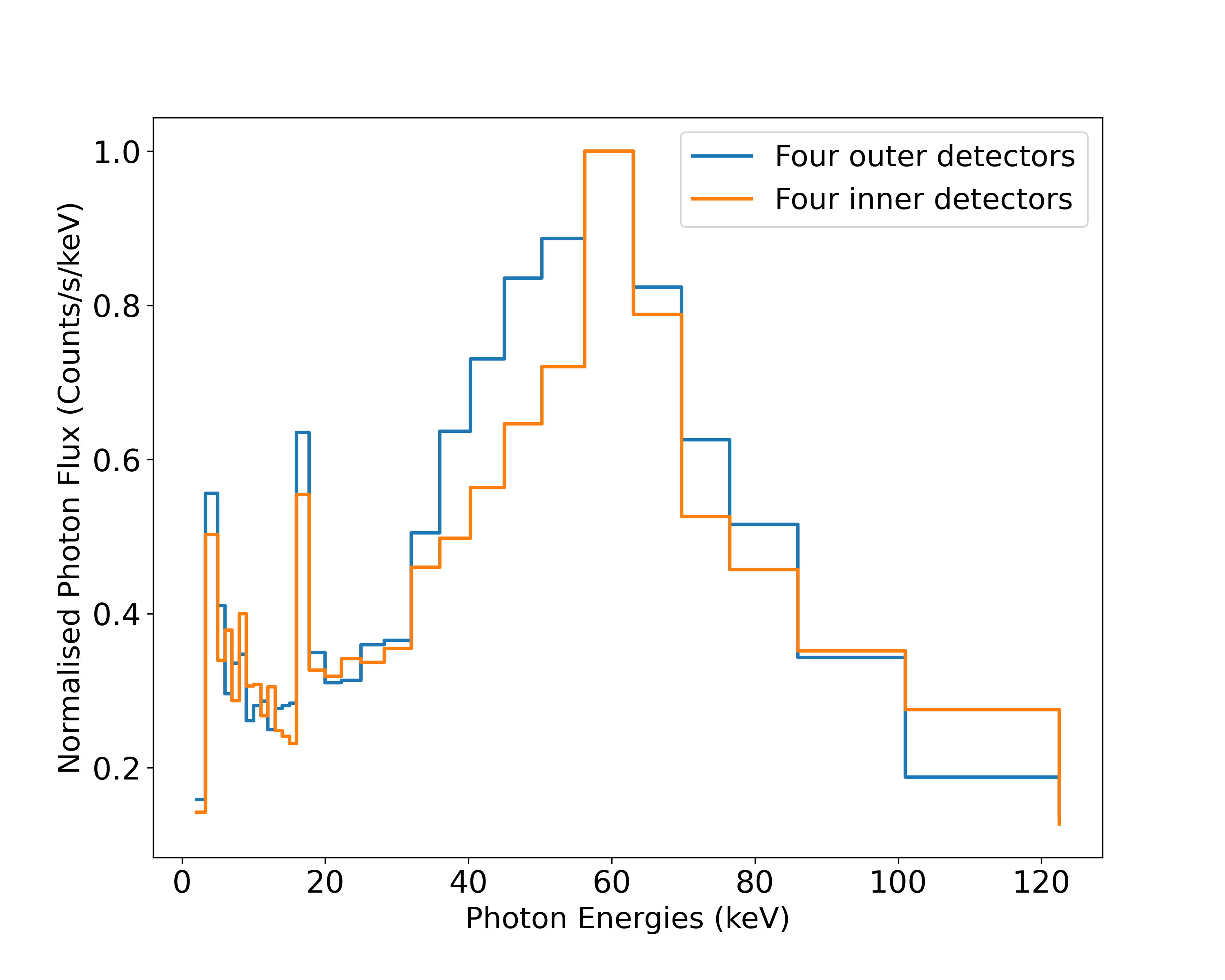}
    \caption{Background subtracted normalised spectrum of HXR photons detected by STIX during the 25th July IP shock crossing for the four outer (brightest) and four inner (dimmest) detectors (detectors 11, 16, 17, 22 \& 13, 14, 19, 20 in Fig \ref{fig:shock_det_view}, respectively).}
    \label{fig:shock_spectrum}
\end{figure}

\begin{figure}
    \centering
    \includegraphics[width=\columnwidth]{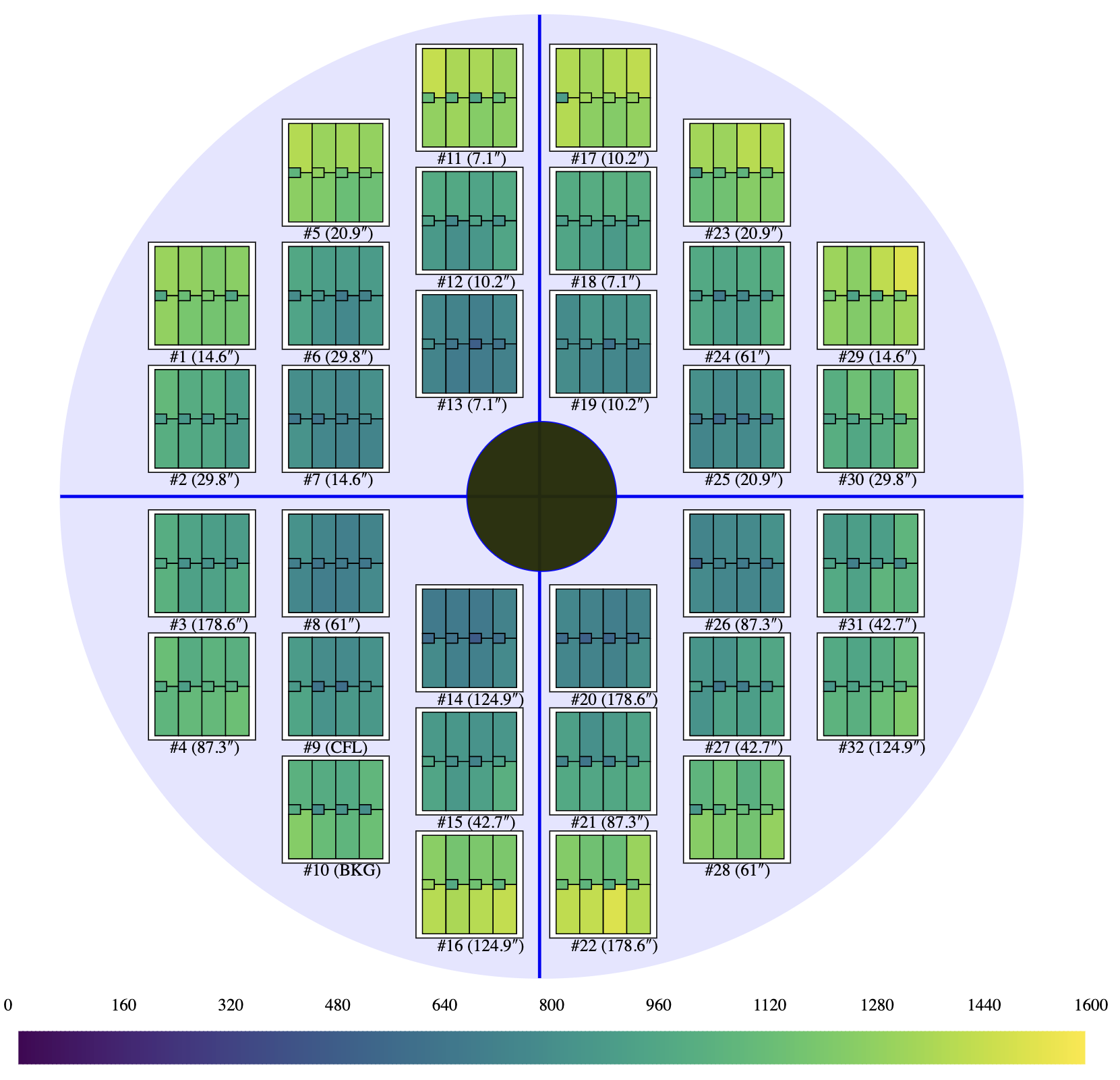}
    \caption{STIX detector view of the 36-76 keV photon counts during the July 25th 2022 shock event. The figure shows a ring-like structure whereby the flux of HXR photons is lower for inner detectors than outer detectors.}
    \label{fig:shock_det_view}
\end{figure}

To understand the mechanism producing the excess HXR counts, the induced HXR spectrum is analyzed. The spectrum for the time interval 2022-07-25 06:24 to 06:31  is shown in Fig. \ref{fig:shock_spectrum}. It has been background subtracted by removing the pre-existing signal in the relevant detectors at an earlier time, namely between 05:36:38 and 05:56:37. The detector pattern of the observed contamination in the energy range 36-76 keV during this interval reveals a ring-like structure (shown in Fig. \ref{fig:shock_det_view}) whereby outer detectors are brighter than inner detectors by a factor of $\sim 1.8$. This pattern is not present during background intervals. Furthermore, there is rotational asymmetry (detectors \#11, 16, 17 \& 22 are brighter than detectors \#2, 3, 30, 31). This indicates that there are geometrical effects contributing to the contamination. 

\subsection{Earth gravity assist maneuver of Solar Orbiter}
On November 27th, 2021 the Solar Orbiter spacecraft performed an Earth gravity assist maneuver. This maneuver required that the spacecraft enter Earth's outer radiation belts, and as a result, the on board instrument suite was switched off for protection. However, for a short duration upon entering the radiation belts, STIX was still on. This resulted in HXR photon counts in the instrument.

\begin{figure}
    \centering
    \includegraphics[width=\columnwidth]{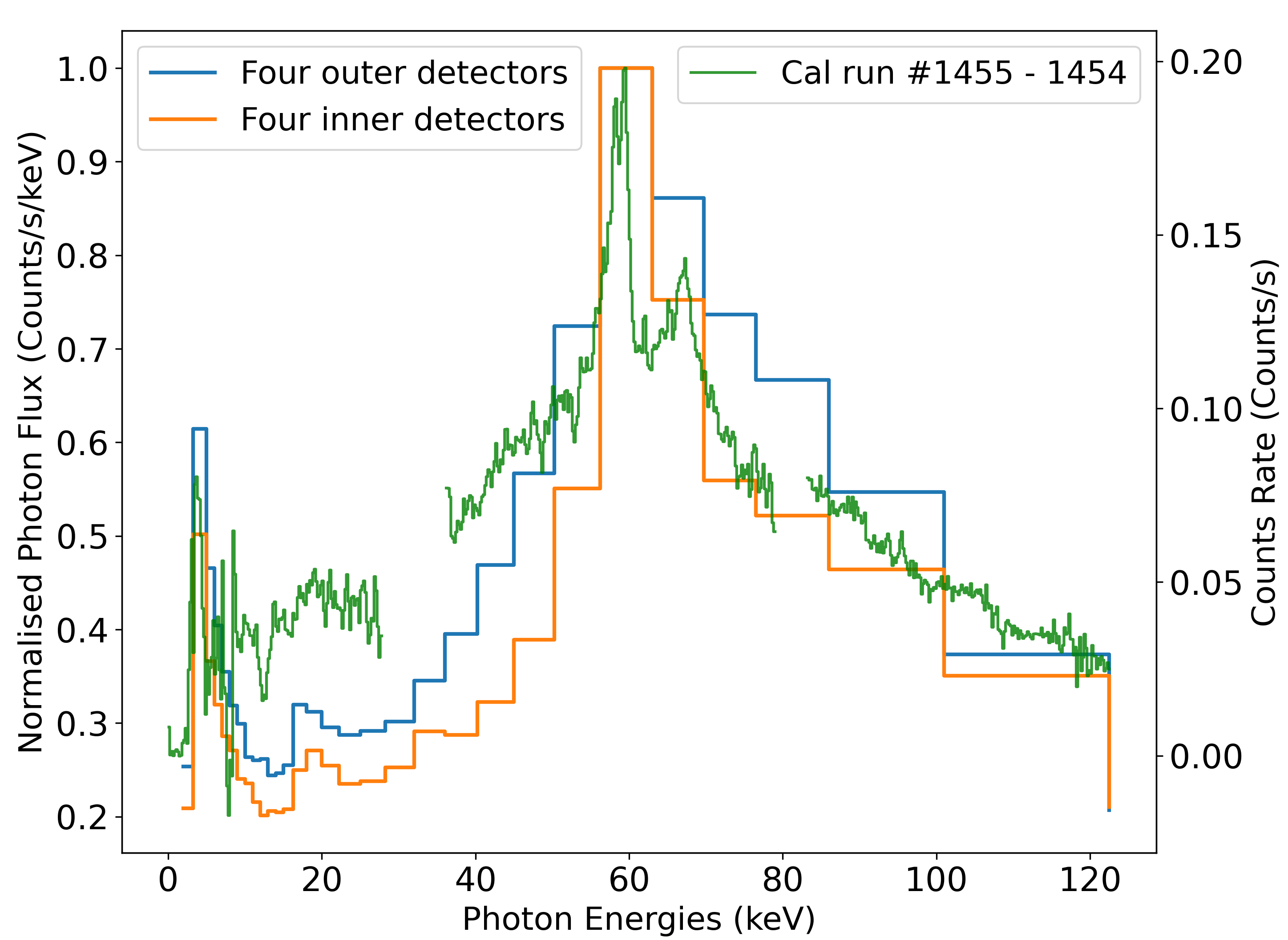}
    \caption{Background subtracted spectrum measured by STIX during the EGAM. The spectra shown are for the four outer (brightest) detectors and the four inner (dimmest) detectors. The overall spectra are similar, though the peak around 60 keV is slightly broader for the outer detectors. This figure also shows the background subtracted calibration spectrum during the EGAM. Gaps in the data are present at strong Ba-133 calibration lines. The observed spectrum includes characteristic line emission of tungsten including the $K_\alpha$ and $K_\beta$ lines and a continuous component.}
    \label{fig:EGAM_spectrum}    
\end{figure}

The observed background subtracted spectrum of HXR photons for the four outer and four inner detectors for the time range 2021-11-27 03:46:09-03:47:52 during the maneuver is shown in Fig \ref{fig:EGAM_spectrum}. An earlier time interval between 2021-11-27 03:10:43 and 03:27:13 was used for background subtraction. The induced spectrum is remarkably similar to the first event shown in Fig. \ref{fig:shock_spectrum}. 

STIX takes a daily calibration spectrum using  128 ${}^{133}$Ba radioactive sources on board and from this higher spectral resolution can be obtained \cite{Krucker_2020}. The calibration spectrum measured during the EGAM is shown in Fig. \ref{fig:EGAM_spectrum}. The calibration spectrum from the previous day has been subtracted from the EGAM calibration spectrum to extract the effect of energetic particles. It shows both characteristic emission lines of tungsten and a Bremsstrahlung component. Since the energy resolution of the calibration spectrum is much higher than the science energy channels, we can resolve distinct line emission of tungsten (e.g. at 57.98, 59.32 \& 67.24 keV) which are spectrally integrated in the other two spectra shown in Fig. \ref{fig:EGAM_spectrum}. 

\begin{figure}
    \centering
    \includegraphics[width=\columnwidth]{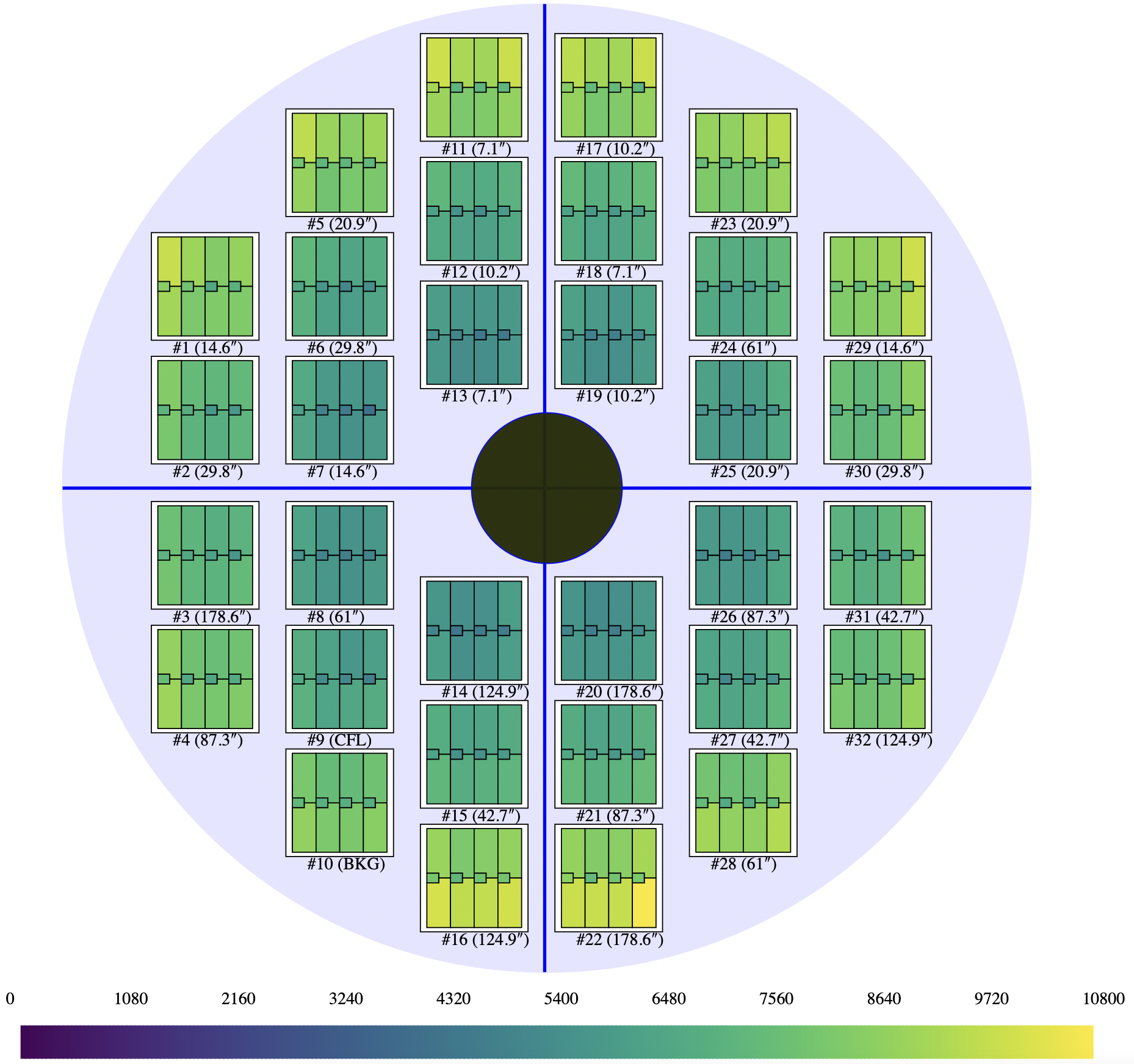}
    \caption{STIX detector view for counts in the energy range 36-76 keV during the EGAM. The counts form a ring-like structure, where the outer ring is brighter than the inner ring.}
    \label{fig:36-76-ring}
\end{figure}

Fig. \ref{fig:36-76-ring} shows the spatial distribution of counts on the STIX detectors with energies in the range 36-76 keV. Similarly to the IP shock crossing event (Fig. \ref{fig:shock_det_view}), the distribution of counts has a ring-like structure.

\section{Modelling}
To gain a deeper understanding of the particle species responsible for the excess signal in STIX and the mechanism by which they produce this signal, modelling efforts were made. In particular, the instrument response to energetic electrons was simulated using the Geant4 software package. Ideally, one would perform a Geant4 simulation which considers the entire spacecraft response, however, since a detailed spacecraft model was unavailable, we employ alternative simplistic analysis methods detailed in the subsections which follow. 

\subsection{Geant4 Simulations of the STIX Instrument Response}
\begin{figure}
    \centering
    \includegraphics[width=\columnwidth]{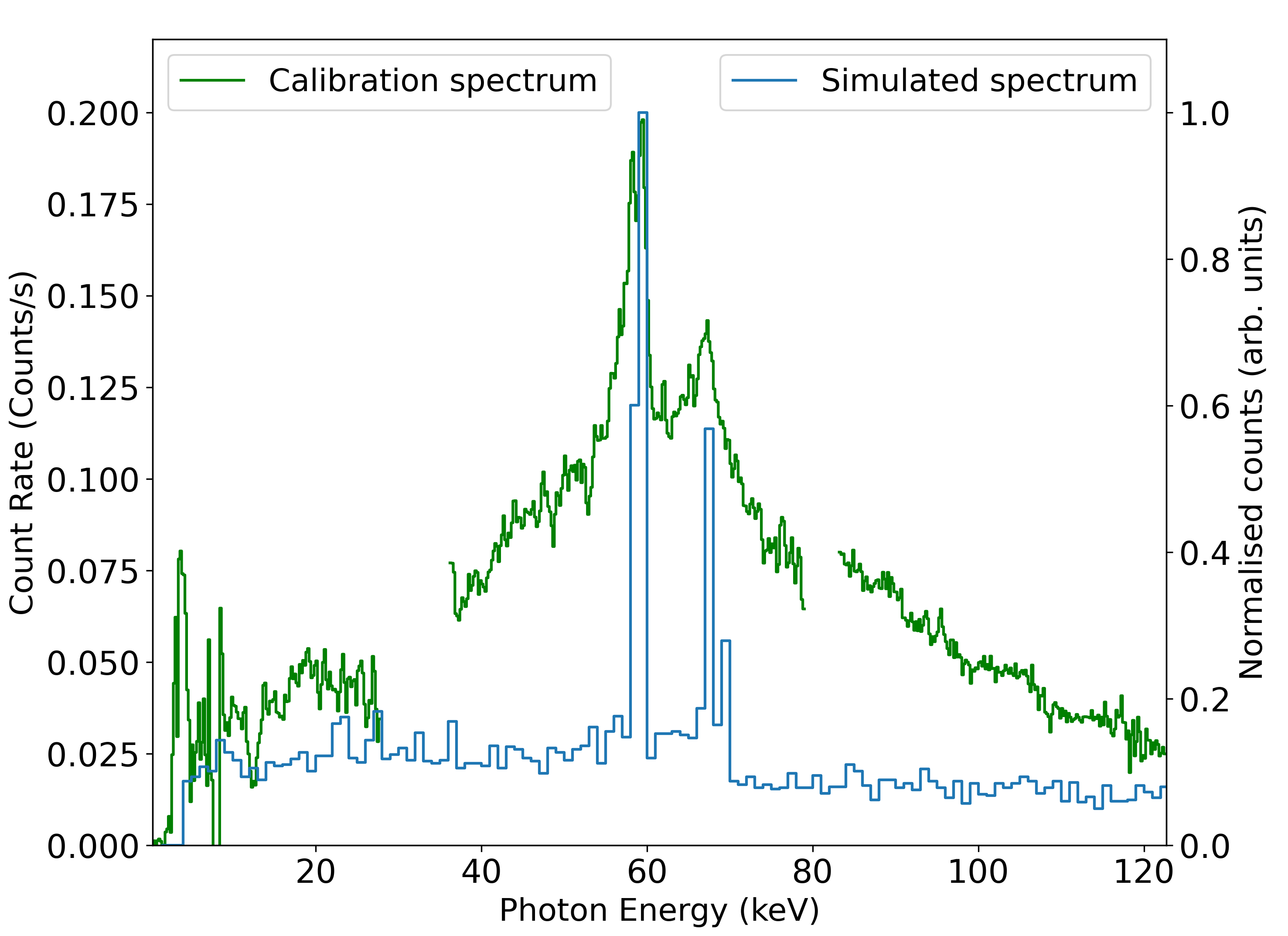}
    \caption{Total simulated spectrum resulting from a normally incident beam of 0.08-8 MeV electrons compared to the measured EGAM background subtracted calibration spectrum. The main tungsten fluorescence emission lines are reproduced, however, the continuous component is not reproduced by the Geant4 simulation.}   
    \label{fig:geant_sim_spectrum}
\end{figure}

Firstly, the STIX instrument response to a flat spectrum of electrons incident normal to the entrance window with energies in the range 0.08-8 MeV was simulated using the Geant4v10.5.0 software package \cite{GEANT42003, ALLISON2016186}. Geant4 simulates many physical processes, the most relevant of which include Bremsstrahlung, Compton scattering, and the photoelectric effect. In this work, the Livermore electromagnetic models were used since the focus here is on low energy processes. Due to the lack of a detailed spacecraft CAD model, the Geant4 simulation does not take into account the spacecraft housing, only the STIX instrument itself. The components in the instrument model include the beryllium entrance windows, the front and rear tungsten grid, 32 Caliste-SO CdTe detectors \cite{caliste}, and the detector electronics module. Fig. \ref{fig:geant_sim_spectrum} shows the resultant photon spectrum measured by the Caliste-SO detectors. In this run, counting statistics are relatively low, in particular, few counts are generated for electron energies $< 3$ MeV. The features that are distinctly present, are the W $K_{\alpha}$ and $K_{\beta}$ lines measured during SEP events. However, the continuous component and the observed ring-like distribution of counts are not reproduced. These results motivate the study of the interaction of energetic electrons with spacecraft components. 

\subsection{AE9 Model of Earth's Radiation Belts Along Solar Orbiter's Trajectory}\label{subsec:omere}
\begin{figure}
    \centering
    \includegraphics[width=\columnwidth]{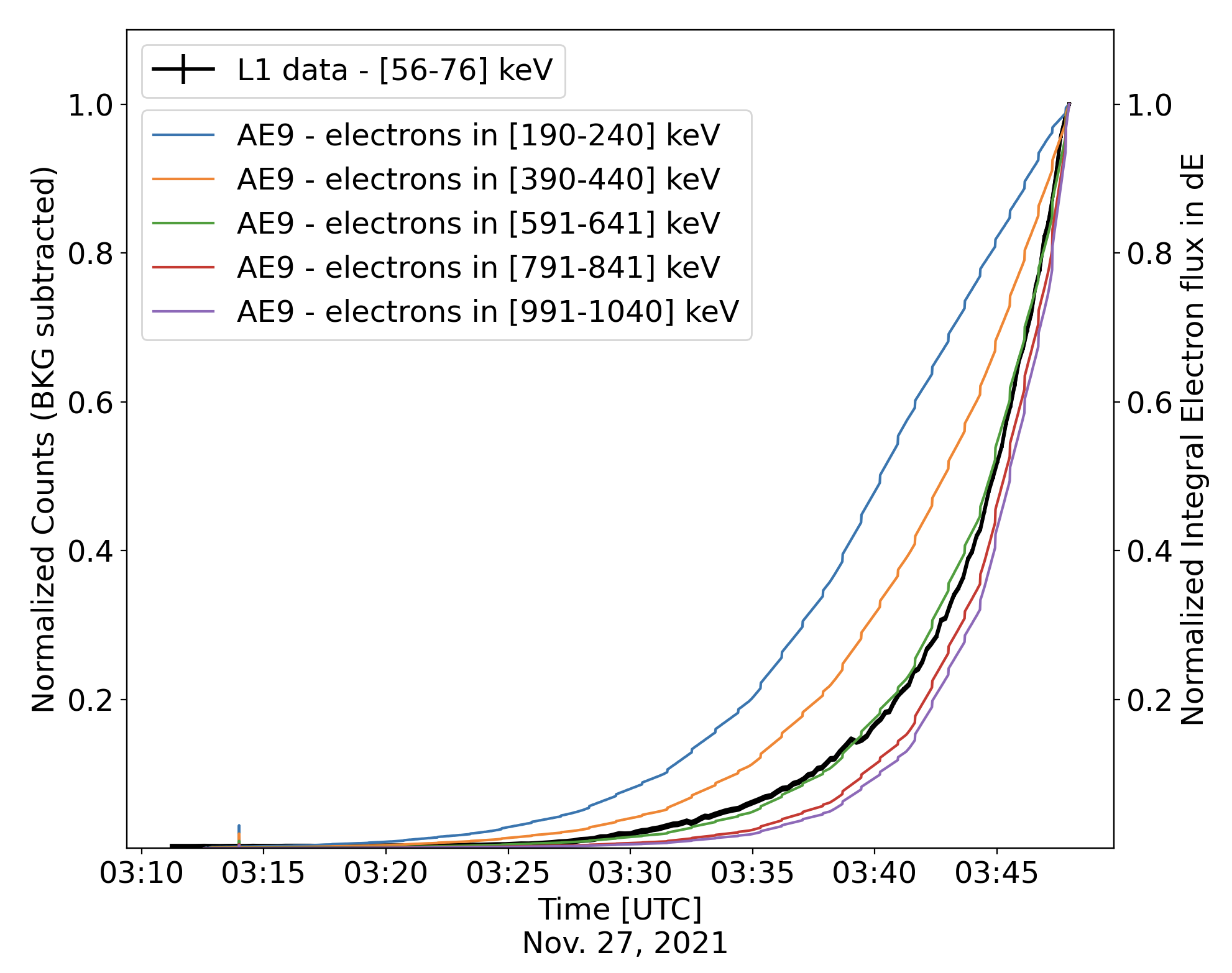}
    \caption{AE9 predicted electron flux along the trajectory of Solar Orbiter during the EGAM compared to 56-76 keV HXR count rate. The electron time profiles across a wide energy range and the 56-76 keV HXR light curve are strongly correlated, indicating that electrons in this energy range could be responsible for the excess HXR counts.}
    \label{fig:predicted_electron_flux_EGAM_OMERE}
\end{figure}

In order to model the expected response during the EGAM, we need to know the particle environment encountered by the satellite while the STIX detector was still switched on up to a low altitude of 17,300 km, where electrons, not protons, are mainly encountered. Since the EPD instrument suite was switched off during the EGAM, the particle environment was estimated based on the most recent in situ measurements using the AE9 \cite{AE9} model available in the OMERE software tool \cite{OMERE}. In practice, we extracted the so-called ‘mean’ (not perturbed) differential spectra of the electrons along the trajectory (obtained from the Solar Orbiter SPICE (Spacecraft, Planet, Instrument, C-matrix, Events) kernels with a time resolution of 1 s and a spectral resolution of 50 keV between 0.04-10 MeV. Electron time profiles across a wide range of energies ($\sim$190-1000 keV) and the 56-76 keV HXR signal are strongly correlated, as shown in Fig. \ref{fig:predicted_electron_flux_EGAM_OMERE}. This indicates that electrons in this energy range could be responsible for the signal detected in STIX. However, this correlation is highly dependent on the radiation belt model used which has certain limitations and uncertainties. Thus, the exact energy range of electrons which correlate best cannot be confidently derived from this analysis.

\subsection{Spacecraft Model}
The lack of a detailed spacecraft model means that we cannot model the spacecraft response to energetic particles directly. However, in the case of the EGAM event, a calibration spectrum with high energy resolution is available. We make use of this information by analytically modelling the Bremsstrahlung and line emission from various potential spacecraft materials. The materials included are based on the limited information available regarding instrument surroundings. The equivalent thicknesses of the materials is tuned by fitting the calculated emission to the calibration spectrum. 

\begin{figure*}
    \centering
    \includegraphics[width=\textwidth]{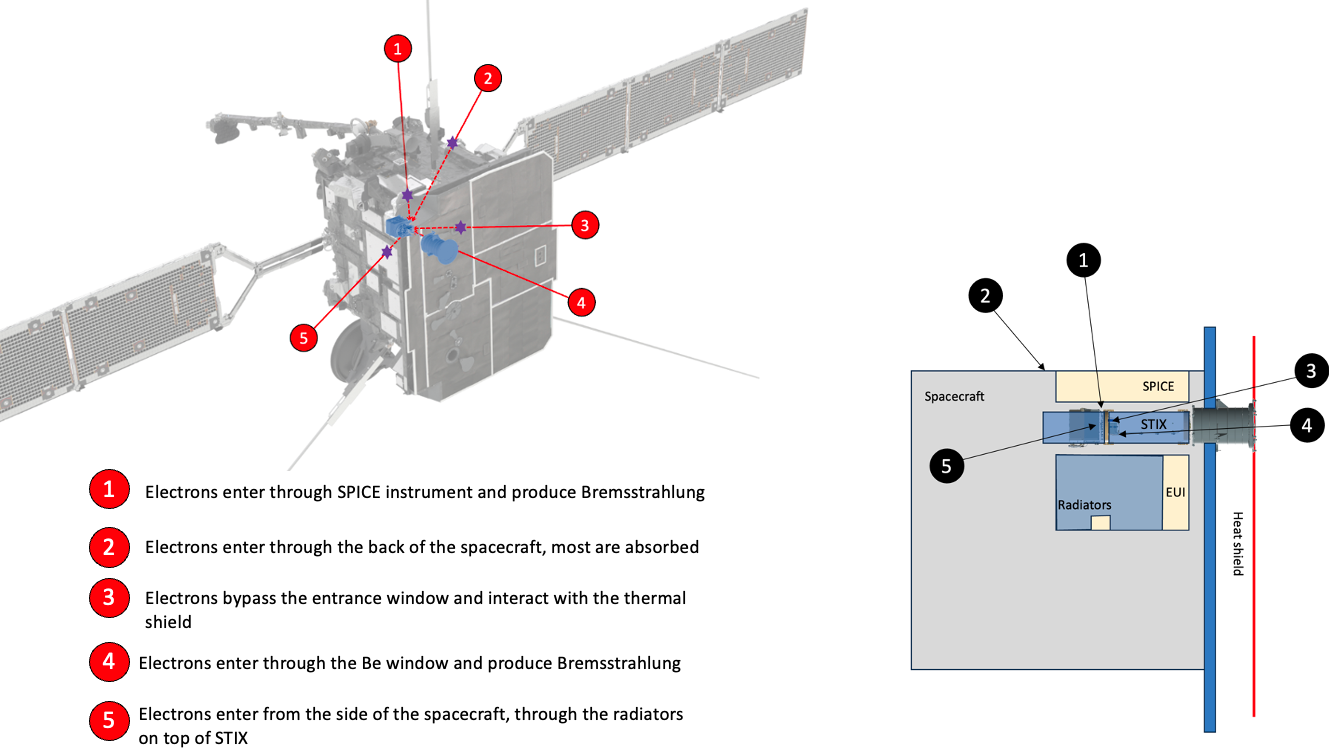}
    \caption{Position of STIX on board Solar Orbiter and a selection of possible electron trajectories traversing the spacecraft to reach STIX. The solid red line indicates the electron trajectories before reaching the spacecraft and the dashed red line indicates the trajectory after reaching and interacting with the corresponding material. The inset on the right-hand side shows a top-down view of STIX, its neighbouring instruments, EUI and SPICE, the heat shield, radiators and the corresponding electron trajectories.}
    \label{fig:spacecraft_housing}
\end{figure*}

The STIX instrument is positioned in Solar Orbiter with the Extreme Ultraviolet Imager (EUI) and The Spectral Imaging of the Coronal Environment (SPICE) instruments on either side. In the front, there is a heat shield which protects from intense solar flux \cite{muller2020}. In addition, in all directions, there are various materials through which particles would need to traverse in order to reach STIX. The left-hand side of Fig. \ref{fig:spacecraft_housing} shows a schematic of STIX in the spacecraft and a selection of paths that particles could traverse to reach STIX. Of course, particles likely enter from all directions. The inset on the right-hand side of Fig. \ref{fig:spacecraft_housing} shows a top-down view of STIX and its surrounding instruments, EUI \& SPICE, and the corresponding particle paths. Specifically, in front of the STIX grids there are two beryllium windows with a combined thickness of 3 mm. In front of the detector electronics module there is a pair of tungsten grids. It is also known that titanium is present in the heat shield and a sandwich panel containing an aluminum alloy \cite{thermal_design}. 

\begin{figure*}
    \centering
    \includegraphics[width=\textwidth]{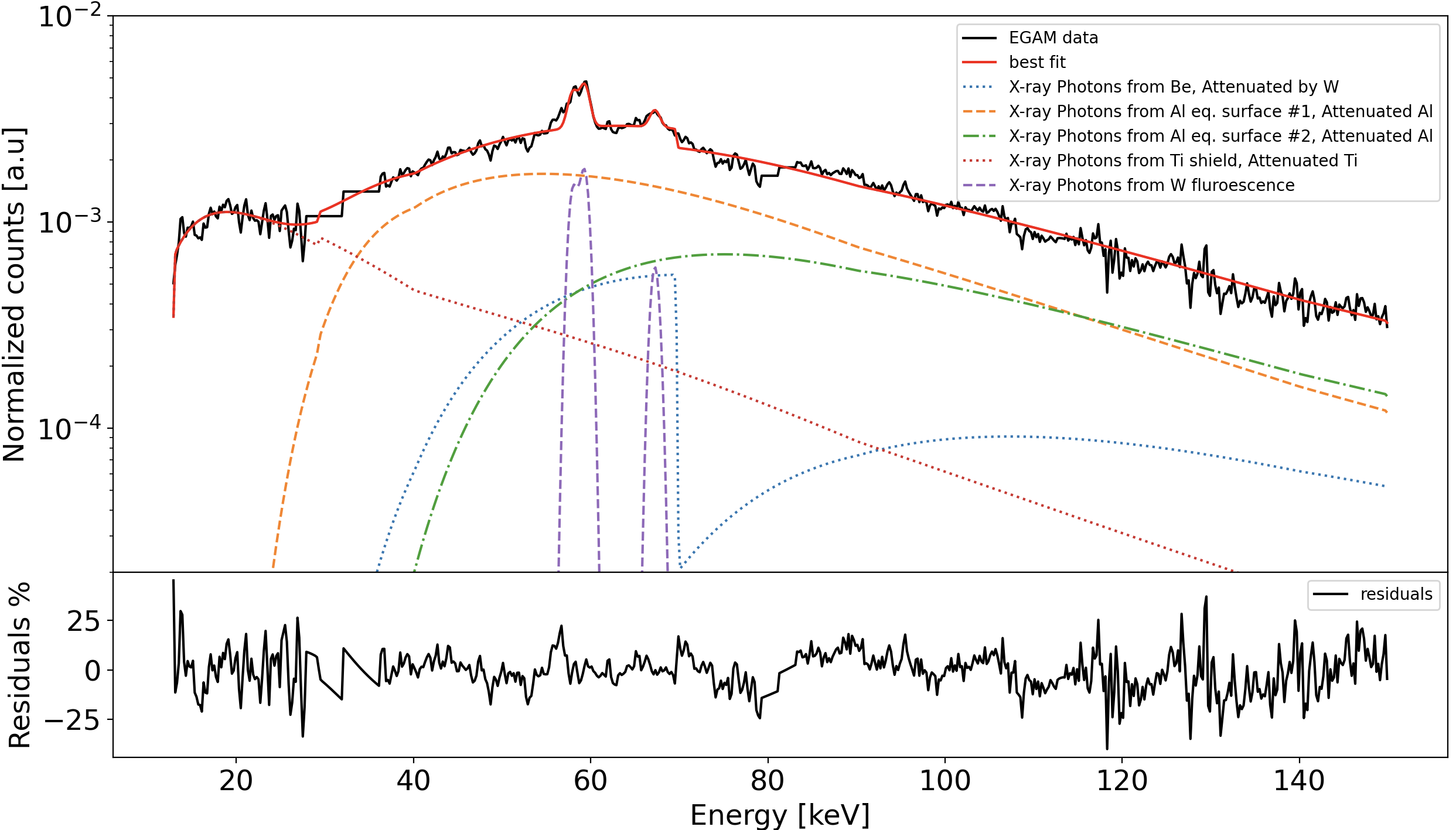}
    \caption{Measured calibration spectrum during the EGAM of Solar Orbiter on 2022-11-27 and the fitted Bremsstrahlung and tungsten fluorescence components. The fitted components include a beryllium target with absorption in tungsten, a thick and thin aluminum target with aluminum absorption, a titanium target with titanium absorption and X-ray fluorescence from the tungsten target. The components fit the data to within $\pm 20$\%. This gives a reduced chi-square statistic $\chi_{\nu}^2 = 0.54$.}
    \label{fig:fit_to_cal_spec}
\end{figure*}

Using this information plus the electron spectrum calculated by the AE9 model (described in subsection \ref{subsec:omere}), the Bremsstrahlung X-ray spectrum produced by the electrons incident on each material was computed. The cross section of electron interaction with atom of atomic number, Z, was computed based on Kramer's model \cite{10.1080/14786442308565244}. The radiation yield (fraction of kinetic energy radiated) was calculated using the NIST ESTAR database~\cite{ESTAR, SELTZER198595, PRATT1977175} for three materials: aluminum, titanium, and beryllium. Absorption of photons in the irradiating material was also accounted for using the xraylib package \cite{xraylib}. For the beryllium window, attenuation in the tungsten grids was considered. In the case of the aluminum and titanium surrounding materials, absorption in further aluminum and titanium respectively was calculated. A term accounting for the K X-ray fluorescence lines of tungsten, falling in the energy range of interest, is computed using the xraylib package. The total radiated flux was also calculated as a weighted sum of the different material contributions. The detector response is applied. This models the photo-electric effect but does not consider Comptonisation which is a reasonable approximation for photon energies below 150 keV. The detected spectrum is then convolved with the detector energy resolution to give the predicted spectrum. The calculated spectrum is fit to the data by tuning the equivalent thicknesses of the materials considered and their relative contributions. The equivalent thickness estimates correspond to an average absorption depth of the material. A good fit to the data (within $\pm 20 $\% with a reduced chi-square value of $\chi_{\nu}^2 = 0.54$) is shown in Fig. \ref{fig:fit_to_cal_spec}. For this particular solution, the fit involves:
\begin{enumerate}
    \item The beryllium entrance window with attenuation in tungsten grids of $200 \pm 34$ ${\mu}m$ equivalent thickness. The sharp attenuation at $69.5$ keV is from the W k-edge. 
    \item Two layers of aluminum with equivalent thicknesses of $1.26\pm0.07$ cm and $4.92 \pm 0.42$ cm respectively. 
    \item A titanium surface with equivalent thickness of $62\pm3 $ ${\mu}m$. 
\end{enumerate}
A good fit is obtained with a minimum of 4 Bremsstrahlung components but is not a unique solution since the component list is not necessarily exhaustive. 

The true known total thickness of the tungsten grids is $\sim 800 {\mu}m$ \cite{Krucker_2020}, which is significantly greater than the value given by the model. However, the tungsten sheet modelled here does not include the slits and slats that exist in reality. This could account for the discrepancy between the required thickness for the fit and the true grid thickness. The tungsten grid attenuation is required to obtain a good fit at high energies. The titanium sheet is required in order to fit the bump in the calibration spectrum at low energies ($\lessapprox 20$ keV) and cannot be accounted for by Comptonisation. The real heat shield titanium component has a thickness of $\sim 250$ ${\mu}m$. It is made up of two stacks, the first at the high temperature multi layer insulation (MLI) and the second at the low temperature MLI with a separation of 245 mm \cite{thermal_design}. The second stack is only $25$ ${\mu}m$ thick.  Therefore, it is conceivable that electrons entering at a large angle to the second layer interface with $\sim 60$ ${\mu}m$ of Ti.  

We interpret the thick and thin aluminum components as representing the equivalent thickness of the spacecraft and its components, such as the radiators and other instruments, respectively. From this analysis we can simply infer that the continuum component in the observed EGAM calibration spectrum originates from Bremsstrahlung emission through energetic electrons interacting with the spacecraft. Nevertheless, without a detailed spacecraft model, it is not possible to reliably infer the surrounding materials and their dimensions. 

\section{Discussion}
The two events presented in this work show similar effects on the STIX instrument on board Solar Orbiter when the spacecraft experiences a high flux of energetic particles. Timing, spectral, and spatial analysis has been performed in order to determine the likely candidate for the contamination in STIX during energetic particle events. Comparison with the timing of measured energetic particle fluxes, where available, indicates a strong correlation between the HXR flux detected by STIX and energetic electrons. Spectral analysis reveals prominent tungsten fluorescence lines above a continuum. Interestingly, the spatial distribution of counts in STIX forms a ring-like pattern. Geant4 simulations of the STIX instrument response to a flat spectrum of electrons with energies in the range 0.08-8 MeV were performed. Furthermore, since energetic particle data were not available for the EGAM event, an AE9 model of the electron population along Solar Orbiter's trajectory through Earth's radiation belts was calculated. This was used to fit the calibration spectrum with analytical models of the Bremsstrahlung emission from a combination of the surrounding materials with approximate equivalent thicknesses along with tungsten fluorescence. The estimation of electron fluxes from the AE9 model has limitations including accurately representing variation of radiation belt populations on short timescales. Additionally, the spacecraft model derived from the fit shown in Fig. \ref{fig:fit_to_cal_spec} is overly simplified and is simply an approximation used to demonstrate that the continuous spectral component can be obtained through Bremsstrahlung originating from electron interaction with spacecraft materials. In order to truly understand the instrument's response to energetic particles, a detailed spacecraft model is required.

\begin{figure}
    \centering
    \includegraphics[width=\columnwidth]{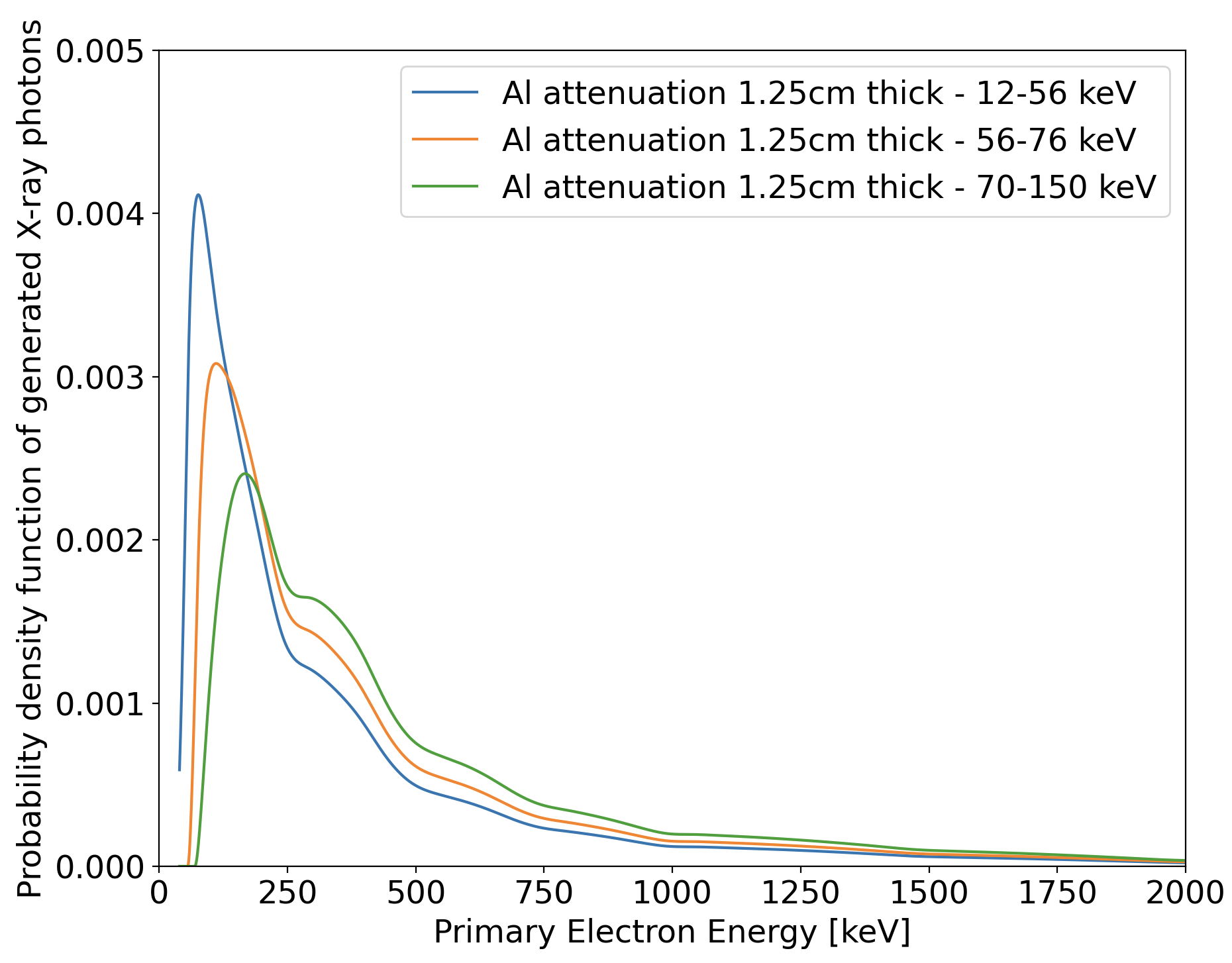}
    \caption{Intensity probability density functions for X-rays generated in 1.25 cm of aluminium in various energy ranges against primary electron energy. For 56-76 keV photons, $\sim 100-700$ keV electrons are the primary source of contamination.}
    \label{fig:relative_photon_int_against_primary_el_en}
\end{figure}

From the derived spacecraft components, we calculated the expected Bremsstrahlung emission for the EGAM event, accounting for attenuation and detector efficiency. The relative intensity of X-ray photons in various energy bands against primary electron energy for the thin aluminum component (1.25 cm) is shown in Fig. \ref{fig:relative_photon_int_against_primary_el_en}. For photons in the main energy range of interest (56-76 keV), electrons in the range $\sim 100-700$ keV are the primary source of the continuum emission. Fig. \ref{fig:relative_photon_int_against_primary_el_en} shows that slightly higher energy electrons produce 70-150 keV photons (above the W K-edge).

\begin{figure}
    \centering
    \includegraphics[width=\columnwidth]{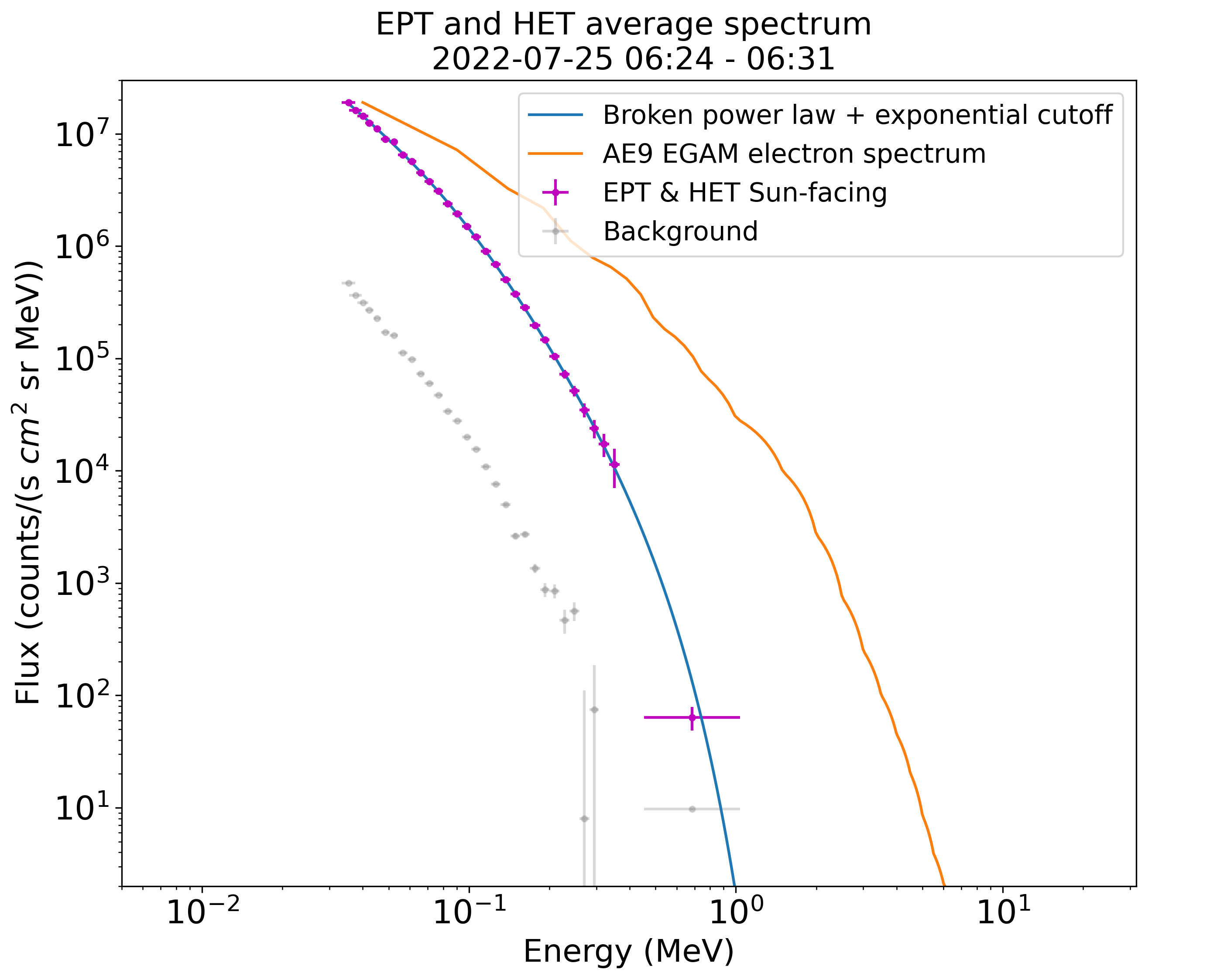}
    \caption{Measured EPD spectrum and fit during the IP shock versus the AE9 predicted EGAM electron spectrum. The broken power law + exponential cutoff fit gives a reduced chi-square, $\chi_{\nu}^2 = 0.06$. The EGAM spectrum is one to three orders of magnitude brighter at $\sim 0.1 - 1$ MeV.}
    \label{fig:EPT_spectrum_vs_egam}
\end{figure}

For the case of the IP shock event, the shock signal was not strong enough in the STIX calibration spectrum to perform the same analysis. We understand this to be due to differences in the electron environment, namely, the electron flux in the relevant energy range ($\sim 0.1 - 1$ MeV) during the shock was one to three orders of magnitude fainter than that in the radiation belt environment. This is shown in Fig. \ref{fig:EPT_spectrum_vs_egam} which shows the electron spectrum measured by EPD for the IP shock event in the time range 06:24-06:31, alongside the AE9 predicted EGAM electron spectrum. The ion contamination correction detailed in section \ref{subsec:IP_shock_obs} has been applied to the electron data prior to fitting. 

\section{Conclusions}
This work demonstrates that electrons with energies $> 100$ keV contribute significantly to excess counts in the HXR signal measured by STIX during energetic particle events. Electrons with energies of approximately 100-700 keV are most efficient at producing this contamination. These electrons produce excess counts from X-ray Bremsstrahlung upon interaction with the materials surrounding the instrument which in turn leads to tungsten fluorescence lines and a Bremsstrahlung component in the observed spectrum. We speculate that the observed ring-like structure of counts falling on the detectors is due to geometrical differences in the surrounding materials and that the rotational asymmetry is due to differences in the material properties of the spacecraft interface and the instruments either side of STIX. We also find that the Caliste-SO detectors are resilient to highly energetic particle events. 

\appendices

\end{document}